\documentclass[epjST]{svjour}
\usepackage{graphics}
\begin{document}
%
\title{Cluster dynamics and cluster size distributions in systems of self-propelled particles}


\author{Fernando Peruani \inst{1} \and Lutz Schimansky-Geier \inst{2} \and Markus B\"ar \inst{3}}
\institute{Max Planck Institute for Physics of Complex Systems, N\"othnitzer Str. 38, 01187 Dresden, Germany \\ 
\and Institute of Physics, Humboldt University Berlin, Newtonstr. 15, 12489 Berlin, Germany\\ 
\and Physikalisch-Technische Bundesanstalt, Abbestr. 2-12, 10587 Berlin, Germany.}
\abstract{
Systems of self-propelled particles (SPP) interacting by a velocity alignment mechanism in the presence of noise exhibit a rich clustering dynamics. 
It can be argued that clusters are responsible for the distribution of (local) information in these systems. 
%
Here, we investigate the  statistical properties of single clusters in SPP systems, 
like the asymmetric spreading of clusters with respect to their moving
direction. 
In addition, we formulate a Smoluchowski-type kinetic model to 
describe the evolution of the cluster size distribution (CSD). 
This model predicts the emergence of steady-state CSDs in SPP systems. 
We test our theoretical predictions in simulations of SPP with nematic
interactions and find that our simple kinetic model 
reproduces qualitatively the transition to aggregation observed in simulations.
}
\maketitle

Examples of large-scale self-organized patterns in systems of 
self-propelled particles  with short-range interactions are found at all scales, from group of animals~\cite{animals,cavagna10,bhattacharya10} and human crowds~\cite{humans} down to insects~\cite{buhl06,romanczuk09}, bacteria~\cite{zhang10}, or actin  filaments~\cite{schaller10}. 
Such patterns are also found in non-living system like in driven granular media~\cite{narayan07,kudrolli08,kudrolli10,deseigne10}.
Despite the fact that the interaction mechanisms between
individual elements are of a different nature, it is possible to determine
some common requirement to achieve large-scale (spatial) self-organization. 
Particularly important for the emerging macroscopic patterns are the self-propulsion of the agents, and their velocity alignment mechanism. 
Such velocity alignment mechanism is often the result of purely physical forces which give rise to an effective alignment interaction as observed in systems of 
self-propelled rods interacting by volume exclusion interactions~\cite{kudrolli08,peruani06,peruani08,baskaran08a,baskaran08b}. 
In other cases, the alignment mechanism emerges from the  complex behavior of the moving entities like in birds~\cite{cavagna10}. 
Simple individual-based models like the Vicsek model~\cite{vicsek95} have helped to reveal the relevance of these two elements, self-propulsion and alignment, by  reducing the problem to the competition between a local aligning interaction 
and some noise~\cite{chate08a}. 
In two dimensions, self-propelled particles moving at constant speed with a ferromagnetic-like velocity alignment   
exhibit at low noise a phase characterized by true long-range polar order which translates into a net flux of particles~\cite{vicsek95,chate08}. 
This ordered phase exhibits several remarkable features like the spontaneous
formation of elongated high density bands that move at
roughly constant speed in the direction perpendicular to the long axis of the band, and anomalous density
fluctuations for low noise levels~\cite{chate08,mishra10}.
When the alignment is replaced by a nematic velocity alignment, 
particles display a phase
characterized by true long-range nematic order at low noise intensity~\cite{ginelli10}. 
Interestingly, spontaneous density segregation into bands is also observed for these
particles, for both, ordered and disordered
phase~\cite{ginelli10}. 
These bands are formed by particles moving in opposite directions along the
long axis of the band. 
In the ordered phase, bands are wide and straight, while bands in the disordered
phase are thinner and highly fluctuating, leading to a zero average global
nematic order. 
For very low noise, bands dissappear and an ordered phase with anomalous
density fluctuations is observed~\cite{ginelli10}.

The two examples we discussed above, self-propelled particles with
ferromagnetic and nematic velocity alignment, display large-scale high-density patterns and seggregation in
absence of any attracting force~\cite{chate08,ginelli10}. 
Interestingly, this also occurs at densities below the percolation
threshold.
%
%
Clearly, a prerequisite for the emergence of ordered phases is the establishment of
some sort of effective interaction/communication among the particles. 
Such interaction/communication should allow information on the local order to
be distributed across the system in order to reach global order. 
On the other hand, this local information is transported by the particles
themselves as they move. 
At this point it is important to notice that local (velocity) alignment
 leads to local polar order which results in the formation of clusters of
particles that move  in the same direction. 
In other words, information is often rather transported by these clusters than by
isolated particles, as discussed in~\cite{ginelli10}. 
Altogether, the cluster dynamics of self-propelled
particles plays a key role in the formation of globally ordered phases.

Here, we study first the  properties of single clusters and second  propose a simple
kinetic theory that describes the steady state cluster size distribution
of self-propelled particles. Our approach is a modification of the Smoluchowki kinetic equations that were developped to describe the aggregation of colloids \cite{smoluchowski1917}. 
Clustering effects and the emergence of steady state cluster size distributions in self-propelled particle systems were observed 
in the Vicsek model~\cite{huepe04}, self-propelled rods~\cite{peruani06}, sperm-like swimmers~\cite{yang08}, and swimming~\cite{zhang10} and gliding bacteria~\cite{peruani10}. While transient clusters and aggregation were reported in colloidal self-propelled rods~\cite{wensink08} and in a 1d-system of self-propelled particles~\cite{escudero10}. 
Our theoretical approach aims at finding a common mechanism behind these scattered observations.

We start by defining the equation of motion of generic self-propelled
particles in section 1. Then, we focus on the statical properties of single
clusters in section 2, to finally propose a simple set of equation for the
evolution of the cluster size distribution, section 3. A summary of the obtained results is given in section 4.


\section{Evolution equation of self-propelled particles}

We consider point-like particles moving at constant speed in a two
dimensional space, as proposed in~\cite{peruani08}, and assume an over-damped situation such that the
state of particle $i$ at time $t$ is given by  its position $\mathbf{x}_i$ and its
direction of motion $\theta_i$. The evolution of $i$-th particle is governed by
the following equations:
\begin{eqnarray}\label{eq_mot_x}
\dot{\mathbf{x}}_i&=& v_0 \mathbf{v}(\theta_i)  \\
\label{eq_mot_angle} \dot{\theta_i}&=& - \gamma(v_0) \frac{\partial
U}{\partial \theta_i}(\mathbf{x_i},\theta_i) +\tilde{\eta}_{i}(t)
\end{eqnarray}
where $\gamma(v_0) \propto v_0^{-1}$ is a relaxation constant, and $U$ the interaction potential between particles, and hence $\frac{\partial U}{\partial
\theta_i}(\mathbf{x_i},\theta_i)$ defines the velocity alignment
mechanism. The parameter  $v_0$
represents the active speed of the particles, $\mathbf{v}(\theta_i)$ is defined as
$\mathbf{v}(\theta_i)=(\cos(\theta_i), \sin(\theta_i))$, and $\tilde{\eta}_{i}(t)$ is an additive white noise applied to the
direction of motion.
%

In analogy to spin systems, a ferromagnetic velocity alignment
mechanism is given by a potential defined as:
\begin{equation} \label{eq:ferromagneticpotential}
U_{F}(\mathbf{x_i},\theta_i)=-\sum_{\left|\mathbf{x}_{i}-\mathbf{x}_{j}\right|\leq\epsilon}
\cos(\theta_i-\theta_j)
\end{equation}
where $\epsilon$ is the interaction radius of the particles. For the nematic alignment mechanism, the
potential takes the form:
\begin{equation} \label{eq:liquidcrystalpotential}
U_{LC}(\mathbf{x_i},\theta_i)=-\sum_{\left|\mathbf{x}_{i}-\mathbf{x}_{j}\right|\leq\epsilon}
\cos^2(\theta_i-\theta_j) \,.
\end{equation}
One can add a coupling strength coefficient in front of the sum of Eqs.~(\ref{eq:ferromagneticpotential}) and (\ref{eq:liquidcrystalpotential}). We assume that the
coupling strength is absorbed in $\gamma(v_0)$ in Eq. (\ref{eq_mot_angle}).
Notice that the potential given by Eq. (\ref{eq:ferromagneticpotential})
exhibits one minimum, while Eq. (\ref{eq:liquidcrystalpotential}) has two
minima. The latter situation corresponds to particles moving in opposite directions.

In the limiting case of very fast angular relaxation we obtain 
from Eqs. (\ref{eq_mot_x}) and (\ref{eq_mot_angle}) the updating rules:
\begin{eqnarray}\label{motion_pos}
\mathbf{x}_{i}^{t+\Delta t }&=&\mathbf{x}_{i}^{t} +v_0 \mathbf{v}\left(\theta_i^{t} \right)\Delta t \\
\label{motion_vel} \theta_i^{t+\Delta t }
&=&\arg\left(\sum_{\left|\mathbf{x}_{i}^{t}-\mathbf{x}_{j}^{t}\right|\leq\epsilon}\mathbf{f}(\mathbf{v}(\theta_j^{t}),\mathbf{v}(\theta_i^{t}))\right)+\eta_{i}^{t}
\end{eqnarray}
where  $\arg\left(\mathbf{b}\right)$
indicates the angle of a vector $\mathbf{b}$ in polar coordinates, and  
$\eta_{i}^{t}$ is a delta-correlated white noise of strength $\eta$
($\eta_{i}^{t}\epsilon\left[-\frac{\eta}{2},\frac{\eta}{2}\right]$). Given two vectors $\mathbf{a}$ and $\mathbf{b}$, $\mathbf{f}(\mathbf{a},\mathbf{b})$ is defined as follows. 
For ferromagnetic alignment, $\mathbf{f}(\mathbf{a},\mathbf{b})=\mathbf{a}$ and Eqs. (\ref{motion_pos}) and (\ref{motion_vel}) reduce to  the Vicsek model~\cite{vicsek95}.
For nematic alignment, $\mathbf{f}$ takes the form $\mathbf{f}\left(\mathbf{a},\mathbf{b}\right) = sign(\mathbf{a}.\mathbf{b})\,\mathbf{a}$ and Eqs. (\ref{motion_pos}) and (\ref{motion_vel}) define a minimal self-propelled rod model~\cite{peruani08,ginelli10}.
%
%
%
%

\subsection{Order parameters}

Ordered phases can be characterized by the following order parameters. 
The ferromagnetic - i.e., polar - order parameter is defined by:
\begin{eqnarray}\label{eq:orderparam_f}
S^{F} = \left| \frac{1}{N} \sum_{k=1}^{N} \exp (i\, \theta_k)  
\right| \, ,
\end{eqnarray}
where $N$ stands for the total number of particles in the system, and the
direction $\theta_k$ is represented as a phase in the complex plane. 
This definition is equivalent to  $S^{F} = \left| \frac{1}{N} \sum_{k=0}^{N} \mathbf{v}(\theta_k)  
\right|$.
$S^{F}$ takes the
value $1$ when all particle move in the same direction, while in the
disordered phase, i.e., when particles move in any direction with equal
probability, it vanishes. 

On the other hand, the nematic ordered parameter takes the form: 
\begin{eqnarray}\label{eq:orderparam_lc}
S^{LC} = \left| \frac{1}{N} \sum_{k=1}^{N} \exp (i\,2\, \theta_k)  
\right|. 
\end{eqnarray}
Formally, $S^{LC}$ can be derived from the order parameter matrix $Q$ of {\it liquid
crystals} (LC)~\cite{doi}, as the largest eigenvalue (which here we have normalized
such that $S^{LC} \in [0,1]$).
When the system is perfectly nematically ordered, i.e., when particles move in
opposite directions along the same axis, $S^{LC}$ takes the value $1$. 
A genuine nematic phase implies half of the particles moving in one direction,
and the other half in the opposite direction. 
If this symmetry is broken, for instance, by having $3/4$ of the particles
moving in one direction, and $1/4$ in the opposite, $S^{LC}$ will still be
$S^{LC}=1$, but $S^{F}>0$. In summary, a perfectly polarly ordered phase is 
characterized by $S^{F}=S^{LC}=1$, while for genuine nematically ordered
phase, $S^{F}=0$ and $S^{LC}=1$.

\section{Evolution of a single (isolated) cluster}

This section is mainly devoted to the understanding of the evolution of a
single, isolated cluster. 
We consider a situation in which initially all particles are located at the origin and move in direction $+ \widehat{\mathbf{x}}$ with speed $v_0$. 
%
%
Since initially each particle can see all the others, the problem can be described initially by a mean-field. 
According to Eq.(\ref{motion_vel}), and assuming ferromagnetic interactions, particles calculate the same common direction of motion. 
The additive noise $\eta^{t}_{i}$ acts just as a perturbation around the global common direction of motion. 
This average vector can be thought as an external field that guides the particles. 
Thus, particles perform a directed random walk~\cite{peruani07}. 
Since initially each particle can see all the others, all of them calculate the same average direction $\alpha_0$. 
The initial condition in our example is such that all particles move  in direction $+ \widehat{\mathbf{x}}$ at $t=0$, and thus $\alpha_0 = 0$. 
This assumption is only true for $\eta<\pi$, if particles interact by nematic alignment. 
Then, the angular dynamics of the $i$th-particle can be approximated by: 
\begin{eqnarray}\label{eq:OPinSwarm}
\theta^{t+\Delta t}_{i} = \alpha_0 + \eta^{t}_i \, ,
\end{eqnarray}
where again, for our example, $\alpha_0 = 0$. 
This means that the probability of finding a randomly chosen particle pointing
in direction $\theta$ at $t+\Delta t$ is 
\begin{eqnarray}\label{eq:prob_theta}
P(\theta)= \frac{1}{\eta}g(\theta, \alpha_0, \eta)\, ,
\end{eqnarray}
where $g(\theta, \alpha_0, \eta)$ is defined to be $1$ when $\alpha_0-\eta/2 \leq \theta \leq \alpha_0 + \eta/2$, and $0$ otherwise, i.e., $g(\theta, \alpha_0,
\eta)=H(\alpha_0+\eta/2-\theta).H(\eta/2-\alpha_0+\theta)$, where $H(x)$ is a Heaviside function. 
In Eq.~(\ref{eq:OPinSwarm}) we are assuming that the average direction $\alpha_0$ does not change over time, which is not true for a finite cluster. 
Nevertheless, this simplification makes the problem analytically tractable,
and, as we will see, the obtained results will be of interest.  
Under these assumptions, the position of the $i$-particle at time $t_n=n \Delta
t$ - using the discrete time description, i.e., Eqs. (\ref{motion_pos}) and (\ref{motion_vel}) - is given by:
\begin{eqnarray}\label{eq:x_i}
x_i(t_n) &=& \sum_{k=0}^{n} \cos(\theta_i(t_k))  v_0 \Delta t \\
\label{eq:y_i}
y_i(t_n) &=& \sum_{k=0}^{n} \sin(\theta_i(t_k))  v_0 \Delta t \, .
\end{eqnarray}
From Eq.~(\ref{eq:x_i}) and (\ref{eq:y_i}), and using Eq.(\ref{eq:prob_theta}),
it is possible to derive $\langle x(t_n) \rangle$, $\langle y(t_n) \rangle$,
$\langle x^2(t_n) \rangle$, and $\langle y^2(t_n) \rangle$, where $\langle
... \rangle$ denotes an average over all particles and realizations of the
noise. 
For the calculation, it is  important to assume that $\langle \theta_i(t_k)
\theta_j(t_l)  \rangle = \sigma \delta_{k,l} \delta_{i,j} $, with $\sigma$ the
second moment of distribution given by Eq.(\ref{eq:prob_theta}). 
The calculation becomes straightforward after computing $\langle
\cos(\theta_i(t_k)) \cos(\theta_i(t_k))  \rangle$, $\langle
\cos(\theta_i(t_k)) \cos(\theta_i(t_l))  \rangle$, $\langle
\sin(\theta_i(t_k)) \sin(\theta_i(t_k))  \rangle$, and $\langle
\sin(\theta_i(t_k)) \sin(\theta_i(t_l))  \rangle$.
With this at hand, we derive the diffusion coefficients of the spreading
process. 
To gain intuition on the problem, we express the time evolution of the
cloud of particles in terms of the (continuum time)  particle density
$\rho(\mathbf{x},t)$ which obeys the following equation (assuming  propagation of the particles along $+ \widehat{\mathbf{x}}$, and thus $\alpha_0 = 0$):
\begin{eqnarray}\label{eq:polapol_density}
\partial_{t}\rho(\mathbf{x},t) = -  V(\eta, v_0) \partial_{x}\rho(\mathbf{x},t) +
 \mathbf{\bigtriangledown}.\left( {\mathbf{D}(\eta, v_0)}
 \mathbf{\bigtriangledown}\rho(\mathbf{x},t) \right).
\end{eqnarray}
In  the convective term, $V(\eta)$ is the mean projection of the instantaneous velocity of the particles on the $+ \widehat{\mathbf{x}}$
semi-axis. 
%
$V(\eta)$ is by definition  $V(\eta)=v_0 \int_{0}^{2 \pi} d\theta P(\theta)\cos(\theta)$, that  takes the form:
\begin{eqnarray}\label{eq:V}
V(\eta) = v_0 \frac{2 \sin(\eta/2)}{\eta} \,.
\end{eqnarray}
This is nothing else than $\lim_{t_n \to \infty} \langle x(t_n) \rangle /t_n$.  
Notice that in the limit of $\eta \to 0$, $V(\eta) \to 1$, and thus there is a deterministic transport of particles without any diffusion. 
In the limit of $\eta \to 2 \pi$, $V(\eta) \to 0$, and particles experience 
 diffusive motion and no convective flux. 
Now we focus on the diffusive term of Eq.(\ref{eq:polapol_density}), 
where the diffusion matrix $\mathbf{D}$ takes the form: 
\begin{equation} \label{diffusion_matrix}
\mathbf{D}(\eta,v_0) = \left(
\begin{array}{cc}
D_{x}(\eta, v_0)  & 0 \\
0 & D_{y}(\eta, v_0) \\
\end{array} \right).
\end{equation}
with $D_{x}$ and $D_{y}$ defined as:
\begin{eqnarray}\label{eq:diff_elements_x}
D_{x}(\eta,v_0) &=& \left(\frac{1}{2}-
  \left[\frac{\sin(\eta/2)}{\eta/2}\right]^2+\frac{\sin(\eta)}{2\eta}
  \right) v_{0}^{2} \Delta t \\
\label{eq:diff_elements_y}
D_{y}(\eta,v_0) &=& \left(\frac{1}{2}-\frac{\sin(\eta)}{2\eta}
  \right) v_{0}^{2} \Delta t
\end{eqnarray}
These terms are obtained from $\lim_{t_n \to \infty} (\langle x^2(t_n) \rangle - \langle
x(t_n)\rangle^2 )/t$ and  $(\langle y^2(t_n) \rangle - \langle
y(t_n)\rangle^2 )/t_n$. 
Notice that for small noise  $\eta$, $D_{y} > D_{x}$. This implies that the
cluster spreads more in the direction orthogonal to the common moving
direction. 
As mentioned above, particles with ferromagnetic alignment form macroscopic bands that
move in the direction perpendicular to the longest axis of the band. 
Eqs.(\ref{eq:diff_elements_x}) and (\ref{eq:diff_elements_y}) indicate that at
the level of an individual (isolated) cluster a similar situation arises: the
cluster moves in the direction perpendicular to its longest axis.
Notice that $\Delta t$ in Eqs. (\ref{eq:diff_elements_x}) and
(\ref{eq:diff_elements_y}) plays the role of the inverse of the turning rate,
i.e., $\alpha^{-1}$ in Eq. (10) and Eq. (11) in~\cite{peruani07}. 
The description given by Eq. (\ref{eq:polapol_density}) is valid while the cloud of particles remains one coherent giant cluster. 
For longer times, this picture fails and  particle motion looses its coherence.
The interaction among particles is such that  particles move in the same direction as long as they can see each other. 
It cannot, however, prevent the particles from slowly moving apart due to the fluctuations the direction of motion.

\begin{figure}
\centering
\resizebox{0.55\columnwidth}{!}{\includegraphics{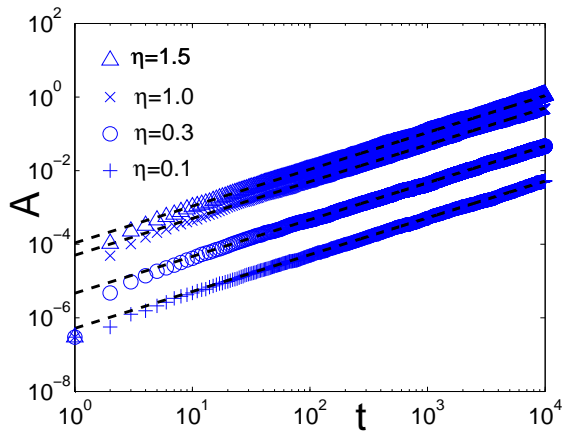}}
\caption{Spreading $A$ of the swarm around it center of mass  vs. time for different values of the noise intensity $\eta$. 
Symbols correspond to measurements of $A(t)$ in simulations with $N=1000$
self-propelled particles with nematic alignment, which initially were located at the origin and all pointing in direction
$+\widehat{\mathbf{x}}$. 
The dashed lines correspond to the approximation given by Eq.(\ref{eq:A})}
\label{fig_diff_msd}
\end{figure}

In the following, we compute the time evolution of the spreading coefficient $A$
of the swarm around its center of mass. 
The spreading area is defined by $A(t_n)=\langle \mathbf{x}(t_n)^2  \rangle - \langle
\mathbf{x}(t_n) \rangle^2$. 
This quantity, $A(t_n)$, can be derived directly from the discrete (time)
process by making use of  Eqs. (\ref{eq:x_i}) and
(\ref{eq:y_i}) as explained above. Alternatively, the continuum time equivalent $A(t)$ can be obtained from
Eq.(\ref{eq:polapol_density}). To obtain  $\langle \mathbf{x}(t) \rangle$, both sides of Eq. (\ref{eq:polapol_density}) are multiplied by $\mathbf{x}$ and integrated over space  to get a simple expression for
$\partial_t\left(\langle \mathbf{x}(t) \rangle\right)$ from which finally  $\langle \mathbf{x}(t) \rangle$ is obtained. 
One proceeds  similarly to get $\langle \mathbf{x}(t)^2  \rangle$, but here both sides
of Eq. (\ref{eq:polapol_density}) are multiplied by $\mathbf{x}^2$. 
The resulting expression is:
\begin{eqnarray}\label{eq:A}
A(t_n) = \left[1-\left(\frac{\sin(\eta/2)}{\eta/2}\right)^2 \right] ( v_{0}^{2}
\Delta t) t_n = D_{eff}(\eta) \, t_n \, .
\end{eqnarray}
A similar expression can be obtained by proposing a continuum-time process for 
the orientation dynamics, where $\Delta t$  is replaced by the inverse of the stochastic turning rate,
see~\cite{peruani07}. 
Notice that in the limit of $\eta \to 0$, $D_{eff}(\eta \to 0)$ vanishes,
while $V(\eta \to 0) = v_0$. Thus, in this limit, the motion becomes purely ballistic.
On the other hand, the diffusive limit corresponds to $\eta \to 2\pi$, where
$V(\eta \to 2\pi) = 0$ and $D_{eff}(\eta \to 2\pi) = v_0^2 \Delta t$. This corresponds to the 
 diffusion coefficient of an ensemble of regular random walkers that  make at each $\Delta t$  
a step with equal probability in any direction.
Fig. \ref{fig_diff_msd} shows a comparison between Eq.(\ref{eq:A}) and simulations 
with $N=1000$ self-propelled particles interacting by  nematic alignment at various noise intensities $\eta$. 
As shown in Fig. \ref{fig_diff_msd}, the agreement between simulations and the predictions of  Eq. (\ref{eq:polapol_density}) is fairly good.

\begin{figure}
\centering
\resizebox{\columnwidth}{!}{\rotatebox{0}{\includegraphics{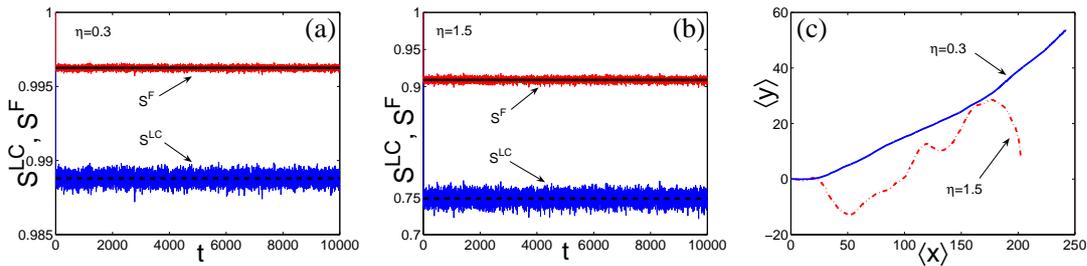}}}
\caption{(a) and (b): $S^{LC}$ and $S^{F}$ vs. time for a  simulation
  performed with $N=1000$ self-propelled particles with nematic alignment, initially located at the origin and all pointing in direction
$+\widehat{\mathbf{x}}$. The noise intensity corresponds to $\eta=0.3$ in (a) and $\eta=1.5$ in (b). The solid black line refers to the prediction for $S^{F}$ given by Eq.
(\ref{eq:sf_perfectorder}),
while the  dashed line is the approximation for $S^{LC}$ given by Eq. (\ref{eq:slc_perfectorder}). 
(c) shows the trajectory of the center of mass of the particle ensemble corresponding to the numerical experiment with $\eta=0.3$ (solid line) and $\eta=1.5$ (dot-dashed line).}
\label{fig_diff_slc_trajectory}
\end{figure}

\begin{figure}
\centering
\resizebox{0.55\columnwidth}{!}{\includegraphics{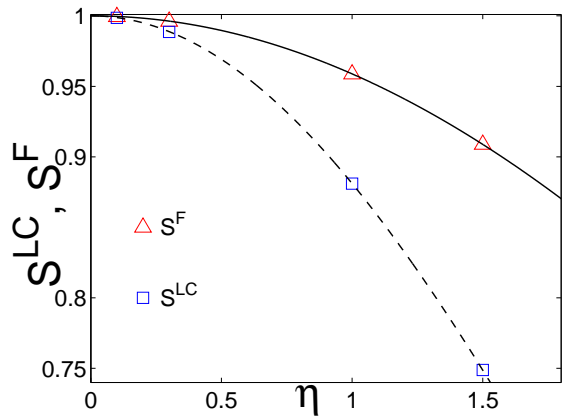}}
\caption{$S^{F}$ and $S^{LC}$ as function of $\eta$ as predicted by Eq. (\ref{eq:sf_perfectorder}) (solid curve) and (\ref{eq:slc_perfectorder}) (dashed curve), respectively. 
Symbols correspond to $S^{F}$ (triangles) and $S^{LC}$  (squares) measured in
simulations performed with $N=1000$ self-propelled particles with nematic alignment initially located at the origin and all pointing in direction
$+\widehat{\mathbf{x}}$.}
\label{fig_diff_slc_eta}
\end{figure}

Now, we turn our attention to the  orientational order parameters exhibited by this cloud of moving particles. 
At first glance, it might seem that, since the bunch of particles moves coherently,  $S^{F}$ and $S^{LC}$ have to be both equal to one.  
Figs. \ref{fig_diff_slc_trajectory}(a) and (b) show $S^{F}$ and $S^{LC}$ as
function of time in simulation with $N=1000$ self-propelled particles with
nematic alignment at two different noise strength values. 
Fig. \ref{fig_diff_slc_trajectory} indicates that $S^{F}$ and $S^{LC}$ inside the coherently moving swarm are still functions of $\eta$. 
We compute $S^{F}$ in terms of $P(\theta)$ by inserting Eq.(\ref{eq:prob_theta}) into the average given by Eq.(\ref{eq:orderparam_f}), and obtain:
\begin{eqnarray}\label{eq:sf_perfectorder}
S^{F}= \frac{2}{\eta} \sin(\eta/2) \,.
\end{eqnarray}
As observed by Dossetti {\it et al.}, this approximation corresponds in fact to the limit of $N \to \infty$ - see Eq. (A8) and its derivation in~\cite{dossetti09}. 
Similarly for $S^{LC}$, by inserting Eq.(\ref{eq:prob_theta}) into Eq.(\ref{eq:orderparam_lc}) we obtain:   
\begin{eqnarray}\label{eq:slc_perfectorder}
S^{LC}= \left|  \left( \frac{\sin(\eta)}{\eta}\right)^2 \right|\,.
\end{eqnarray}
Fig. \ref{fig_diff_slc_eta} compares the predictions of Eqs. (\ref{eq:sf_perfectorder}) and (\ref{eq:slc_perfectorder}) and simulations performed with $N=1000$ particles. 
%
%
The symbols correspond to temporal averages of simulated time series of $S^{F}$ and $S^{LC}$ as shown in Figs. \ref{fig_diff_slc_trajectory}(a) and (b). 
The agreement is remarkably good for the $10^4$ integration steps that the simulations span. 
It is worth to notice that during this period the center of mass of the swarm travels a distance $d$ much larger than the interaction radius $\epsilon$, i.e., $d \gg \epsilon$
(see Fig. \ref{fig_diff_slc_trajectory}(c)),
while the swarm's expansion around its center of mass is comparatively very small. 
Despite the fact that the swarm moves in a quite  coherent way, the value of the order parameters drops with increasing $\eta$ (see Figs. \ref{fig_diff_slc_trajectory}(a) and (b)), and the center of mass performs a more pronounced  meandering
trajectory as shown in Fig. \ref{fig_diff_slc_trajectory}(c). 
Interestingly, the fluctuations of the center of mass around the $\widehat{\mathbf{y}}$ axis are much larger than the swarm spreading around its center. 
This is due to large fluctuations of the average total momentum vector that are not reflected in the fluctuations of  $S^{F}$. 
The total momentum vector fluctuates in direction and modulus, but $S^{F}$ fluctuations are only related to fluctuations of its modulus. 
The same applies to the orientation tensor $Q$ and to its associated  scalar liquid crystal order parameter $S^{LC}$.

\begin{figure}
\centering
\resizebox{0.55\columnwidth}{!}{\includegraphics{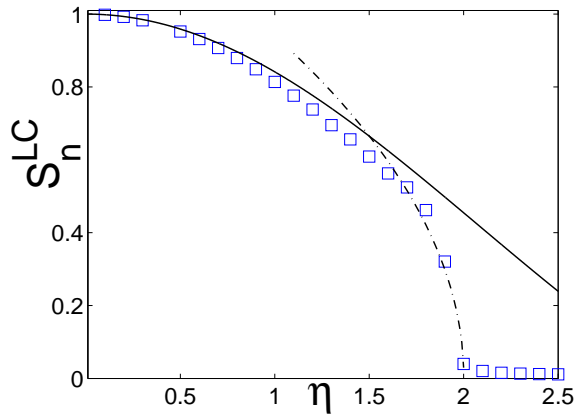}}
\caption{$S^{LC}_{n}$ vs. $\eta$. Symbols correspond to simulations with
  $N=2^{12}$ self-propelled particles with nematic alignment at density $\rho=4$ in a box with periodic boundary conditions and random initial
conditions. 
The dot-dashed curve is a fitting of the first $8$ data points to the left of $\eta_c=2$ through $S^{LC} \sim \left(\eta_c - \eta \right)^{\beta}$, where $\beta = 0.46 \pm 0.03$ (this may not be true for large system sizes).  
The solid curve corresponds to Eq. (\ref{eq:slc_perfectorder}).}
\label{fig_diff_slc_twoapproximations}
\end{figure}

As said above,  this simple description of the swarm is valid until the spreading around the center of mass is such that the density of the moving cluster falls below percolation. 
The critical density is obtained by assuming overlapping discs of radius $r=\epsilon/2=1/2$, $\rho_{p} \sim 1.45$~\cite{quintanilla00}. 
Now, let us imagine that the swarm does not evolve in an infinite space, but
in a box with periodic boundary conditions. Let us assume in addition that the size $L^2$ of the box is such that $N/L^2$ is much larger than $\rho_p$.    
Can this system be described in terms of such simple equations as Eq. (\ref{eq:sf_perfectorder}) and (\ref{eq:slc_perfectorder}) for all times? 
If the initial condition of the system is completely random, does the system reach the same steady state? 
For a random initial condition, we cannot expect a steady state with a unique direction of motion. 
We assume that two main opposite direction of motion emerge, $\alpha_0$ and $\alpha_0+\pi$, and that the angular dynamics of the $i$-th particle is simply given by: 
\begin{eqnarray} \label{function_f2}
\theta^{t+\Delta t}_{i} = \left\{
\begin{array}{lcr}
\alpha_0 + \eta^{t}_{i} & \mbox{with probability } & p_{+} \\
(\alpha_0+\pi) + \eta^{t}_{i} & \mbox{with probability } & p_{-}
\end{array} \right.
\end{eqnarray}
In consequence, $P(\theta)=(p_{+}/\eta)g(\theta,\alpha_0,\eta)+(p_{-}/\eta)g(\theta,\alpha_0+\pi,\eta)$. Using this expression to compute the order parameters $S^{F}$ and $S^{LC}$, we find that $S^{F}=(2/\eta)\sin(\eta/2)(p_{+}-p_{-})$, while $S^{LC}$  is given by Eq. (\ref{eq:slc_perfectorder}). 
This hypothesis is tested in Fig. \ref{fig_diff_slc_twoapproximations}. The simulations were performed with $N=2^{12}$ particles at high density ($\rho=4$) in a box with periodic boundary conditions. The initial condition was random and the simulations were carried out for $10^6$ time steps. 
The solid curve corresponds to the approximation given by Eq. (\ref{eq:slc_perfectorder}), where no fitting parameter is used.
The dot-dashed curve is a fitting of the first $8$ data points to the left of
$\eta_c=2$ assuming $S^{LC} \sim \left(\eta_c - \eta \right)^{\beta}$, where
$\beta=0.46 \pm 0.03$. This assumption may turn to be wrong for large system sizes. 
The figure shows that Eq. (\ref{eq:slc_perfectorder}) provides a good approximation of $S^{LC}$ for small values of $\eta$, but fails to describe  orientational order at large values of the noise intensity, i.e., close to the transition point.  
%
%
%
In summary, Fig.~\ref{fig_diff_slc_twoapproximations} shows that at least for small system sizes  
the ordering dynamics can be described through the above approach for low noise intensity, i.e., far away from the the critical point.  
%
%
The obtained results cannot be used to extrapolate the behavior of the system
to very large system sizes, though well in the ordered phase they may still hold. 
%
This would require a systematic analysis of finite-size effects in large-scale simulations, which is beyond 
 the scope of this work. 
Moreover, we known from~\cite{ginelli10} that in large systems of particles
obeying Eqs.(\ref{motion_pos}) and (\ref{motion_vel}), fluctuations play a crucial role in the
emerging macroscopic dynamics. 
%

\section{Cluster size distribution at low density}

\begin{figure}
\centering
\resizebox{1.0\columnwidth}{!}{\rotatebox{0}{\includegraphics{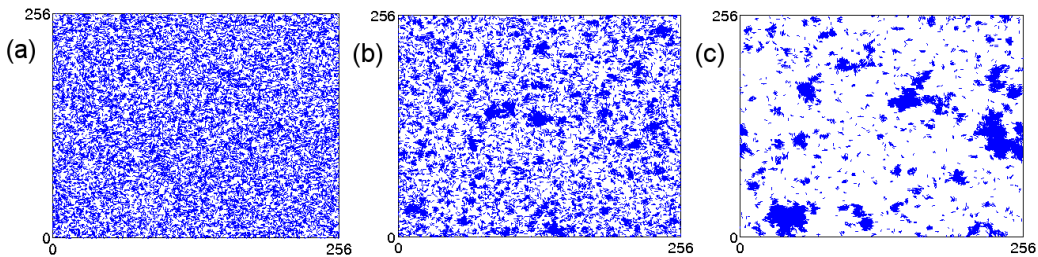}}}
\caption{Snapshots of simulations with self-propelled particles with nematic alignment at density $\rho=0.25$ after $2.5\,10^5$ times units. The values of $\eta$ are: $1.4$ in (a), $1.3$ in (b), and $0.6$ in (c). 
The spatial arrangement of particles seen in the figures is representative of what is observed at the steady state.} \label{fig_snapshots}
\end{figure}

In this section we derive a simple theory to understand the emergence of
steady state cluster size distributions in self-propelled particle systems. 
We illustrate the phenomenon of cluster formation in these systems by performing simulations of
self-propelled particles with nematic alignment. 
Fig. \ref{fig_snapshots} shows simulation snapshots for three values of the
angular noise $\eta$. Notice that as $\eta$ is decreased, clusters become
significantly larger.

Formally, the cluster dynamics of  these systems can be described by deriving
a master equation for the evolution of the probability $p(\mathbf{n}(t))$, where
$\mathbf{n}(t)={n_1(t), n_2(t), ..., n_N(t)}$, with $n_1(t)$ being the number
of isolated particles, $n_2(t)$ the number of two-particle clusters, $n_3(t)$
the number of three-particle clusters, etc. 
This kind of approach has also been used to understand equilibrium nucleation in
gases, where the transition probabilities between states are function of the
associated free energy
change~\cite{schimansky1986,schweitzer1988a,schweitzer1988b}. 
Here, however, we will adopt an alternative more phenomenological strategy for our
non-equilibrium problem. 
Instead of looking for the complete description of the clustering process in terms of
 $p(\mathbf{n}(t))$, we will derive equations for the time evolution of the
 mean value of $n_1(t), n_2(t), ..., n_N(t)$.
To ease the notation, we will refer to $\langle n_1(t) \rangle$,   $\langle
n_2(t) \rangle$, ...,  $\langle n_N(t) \rangle$ simply as $n_1(t)$, $n_2(t)$,
etc.

\begin{figure}
\centering
\resizebox{0.55\columnwidth}{!}{\includegraphics{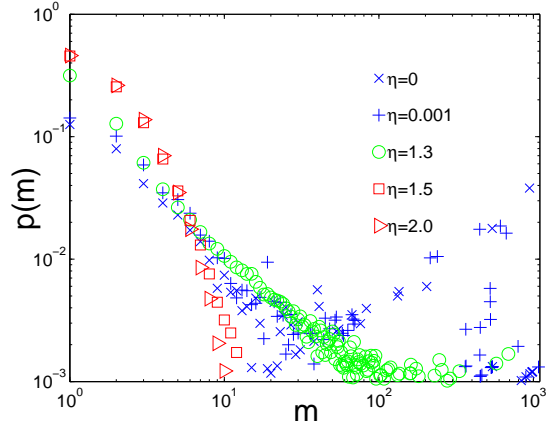}}
\caption{Clustering at different values of the noise intensity $\eta$ at low
  density. Simulation performed with  $N=2^{14}$ self-propelled particles with
  nematic alignment at density $\rho=0.25$. 
Notice the transition from a monotonically decreasing distribution for large
values of $\eta$ to  a non-monotonic distribution with a peak at large cluster
sizes  for small enough values of noise intensity $\eta$.} \label{figs_clustering_d025}
\end{figure}

The simulations were performed with $N=2^{14}$ particles at low density, $\rho=0.25$. We observe a transition from an apparently homogeneous density at large noise values to pronounced formation of clusters at low 
noise (see Fig.~\ref{fig_snapshots}). 
Fig. \ref{figs_clustering_d025} shows the (weighted) cluster size distribution $p(m)=(m\,n_m(t))/N$ for various values of the noise intensity $\eta$. 
We have considered that a cluster is an ensemble of {\it connected} particles. Two
particles are {\it connected} whenever they are separated a distance less or
equal to $d$ -  in Fig.~ \ref{figs_clustering_d025}, $d=2\epsilon$.
The figure indicates that there is transition in the cluster size distribution from a monotonically decreasing
distribution for large values of $\eta$ to a distribution with a peak at large
cluster sizes for small enough noise values. 
Interestingly, at the transition point the cluster size distribution follows a power-law. 

We look for an explanation for the observed clustering phenomena.
%
%
Through the study of the coherence of an initially perfectly oriented swarm, we have learned that clusters have a finite life-time 
before they break into parts. 
Now we incorporate the fact that a moving cluster collects  particles whose relative direction of motion is such that $|\Delta \theta|<\pi/2$, provided the noise intensity $\eta$ is low enough. 
If the topology of the system is a finite torus, these two effects, spreading of particles due to fluctuations in the direction of motion 
and collection of particles due to random {\it collisions} of clusters, reach an equilibrium and the cluster size distribution (CSD) becomes a steady distribution.
As mentioned above, we look for a description of the process in terms of the  number $n_i(t)$ of clusters
with $i$ particles at time $t$. 
The time evolution equations for  the  $n_i(t)$ have the following form:
\begin{eqnarray}
\dot{n}_{1}&=&2B_{2}n_{2}+\sum_{k=3}^{N}B_{k}n_{k}-\sum_{k=1}^{N-1}A_{k,1}n_{k}n_{1}
\nonumber \\
\dot{n}_{j}&=&B_{j+1}n_{j+1}-B_{j}n_{j}-\sum_{k=1}^{N-j}A_{k,j}n_{k}n_{j}
\nonumber \\
&&+\frac{1}{2}\sum_{k=1}^{j-1}A_{k,j-k}n_{k}n_{j-k} \ \quad \mbox{for} \quad j = 2, .....,N-1
\nonumber \\
\dot{n}_{N}&=&-B_{N}n_{N}+\frac{1}{2}\sum_{k=1}^{N-1}A_{k,N-k}n_{k}n_{N-k} \label{rea_vicsek}
\end{eqnarray}
where the dot denotes the time derivative, $B_{j}$ represents the
 rate for a cluster of mass $j$ to loose  a particle, and is defined as 
\begin{eqnarray}\label{eq:rate_splitting}
B_{j}= \frac{D_{eff}(\eta)}{d^2}\sqrt{j} \, ,
\end{eqnarray}
and $A_{j,k}$ is the collision rate
between clusters of mass $j$ and $k$, defined by
\begin{eqnarray}\label{eq:rate_collision}
A_{j,k}= \frac{v_0 2 \epsilon}{a}\left(\sqrt{j}+\sqrt{k}\right) \, ,
\end{eqnarray}
where $a=L^2$ is the area of the two-dimensional space where particles move. 
In Eq. (\ref{eq:rate_splitting}), $d$ denotes, as before, the maximum distance that two
particles can be separated apart to be considered as {\it connected}.
In this way, $d^2/D_{eff}(\eta)$ is the characteristic time a particle
on a cluster boundary needs to detach from a cluster. 
The splitting rate $B_{j}$ is proportional to the inverse of this characteristic time multiplied by the number of particles on the boundary, which we estimate as $\sqrt{j}$. 
On the other hand, the collision rate $A_{j,k}$ is derived in analogy to the collision rate in kinetic gas theory~\cite{reif}  between two  disk-like particles A and B which is known to be proportional to 
the relative velocity of the particles and the sum of their diameters.
We have approximated the diameter of a cluster of mass $j$  by $\epsilon
\sqrt{j}$.

\begin{figure}
\centering
\resizebox{0.55\columnwidth}{!}{\rotatebox{0}{\includegraphics{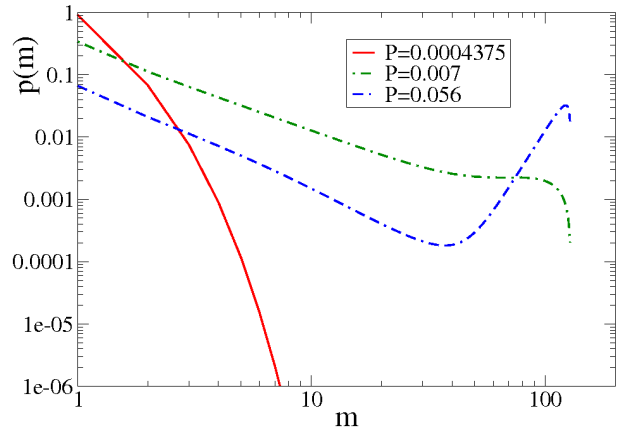}}}
\caption{Steady state cluster size distributions obtained from numerical
  integration of  Eqs.(\ref{rea_vicsek}) with $N=128$ for various values of
  the dimensionless parameter $P$, where $P=\frac{2 \epsilon d^2 v_0}{a D_{eff}(\eta)}$. 
Notice the transition from a monotonically decreasing distribution  for small 
values of $P$ to  a non-monotonic distribution with a peak at large cluster
sizes  for large values of $P$.} \label{fig_csd_theo}
\end{figure}

Direct numerical integration of  these equations shows that Eq.(\ref{rea_vicsek}) produces 
qualitatively similar distributions as the one observed in individual-based
simulations, see  Fig.~\ref{fig_csd_theo}. 
The different curves correspond to various values of the dimensionless
parameter $P$, defined as $P=\frac{2 \epsilon d^2 v_0}{a D_{eff}(\eta)}$. 
For small values of $P$, which correspond to large values of $\eta$, the
distribution $p(m)$  monotonically decreases with $m$, while for large
values of $P$, resp. small values of $\eta$, a peak at large cluster sizes
emerges. 
A quantitative comparison between Eq.(\ref{rea_vicsek}) and individual-based simulations is beyond the scope of this work. 
%
 %
%

\section{Conclusions}

Clusters play a fundamental role in the
macroscopic dynamics of self-propelled particle systems. 
Particularly, clusters are capable to transport orientation information for long distances.
%
Orientational order and cluster dynamics are often closely linked. 

Here we have addressed the problem of cluster dynamics in self-propelled
particle systems in a systematic way. 
We have first studied single isolated clusters and learned that their dynamics
can be understood in terms of simple mean-field arguments. 
We have derived expression for the diffusion coefficients of clusters and
found that the spatial spreading of clusters is anisotropic, with
the long axis of the cluster perpendicular to its moving direction. 
We have also shown  that the mean-field approaches used to characterize
the cluster dynamics can
be applied to describe the behavior of self-propelled particles for  small system sizes far away from the transition point. 

Finally, we have analyzed  the emergence of
steady state cluster size distributions in systems of self-propelled
particles that obey Eqs. (\ref{eq_mot_x}) and (\ref{eq_mot_angle}). 
Particularly, we have derived a set of equations of the Smoluchowski type to
describe the cluster dynamics in the system. 
From direct numerical integration of these equations, we have shown that this 
approach leads to qualitatively similar distributions. 

This study presents  a first step towards the understanding of the
complex clustering behavior exhibited by self-propelled particle systems. 
Several key aspects have been left out in this first approach.  
For instance, a system size analysis is required to check the validity of these
results in large systems. 
Correlations among clusters have been completely ignored, though they may be
crucial to the macroscopic dynamics. 
Finally, we suspect that clustering effects may induce giant number fluctuations as the ones predicted in~\cite{ramaswamy03}. However, a link between these two phenomena has not been still stablished. 
All these issues shall be the subject of future studies.


\begin{thebibliography}{99}

\bibitem{animals}  Three Dimensional Animals Groups, edited by J.K. Parrish and W.M Hamner (Cambridge University Press, Cambridge, England, 1997).

\bibitem{cavagna10} A. Cavagna et al., Proc. Natl. Acad. Sci. 107, 11865 (2010).

\bibitem{bhattacharya10} K. Bhattacharya and T. Vicsek, New J. Phys. 12, 093019 (2010).

\bibitem{humans} D. Helbing, I. Farkas, and T. Vicsek, Nature (London) 407, 487 (2000).

\bibitem{buhl06} J. Buhl et al.,  Science 312, 1402 (2006).

\bibitem{romanczuk09} P. Romanczkuk, I.D. Couzin, and L. Schimansky-Geier, Phy. Rev. Lett. 102, 010602 (2009).

\bibitem{zhang10} H.P. Zhang et al., Proc. Natl. Acad. Sci. 107, 13626 (2010).

\bibitem{schaller10} V. Schaller et al., Nature 467, 73 (2010).



\bibitem{narayan07} V. Narayan, S. Ramaswamy, and N. Menon, Science 317, 105 (2007).


\bibitem{kudrolli08} A. Kudrolli, G. Lumay, D. Volfson, and L.S. Tsimring, Phys. Rev. Lett. 100, 058001 (2008).

\bibitem{kudrolli10} A. Kudrolli, Phys. Rev. Lett. 104, 088001 (2010). 


\bibitem{deseigne10} J. Deseigne, O. Dauchot, and H. Chat\'e, Phys. Rev. Lett. 105, 098001 (2010). 


\bibitem{peruani06} F. Peruani, A. Deutsch, and M. B{\"a}r, Phys. Rev. E, 74, 030904 (2006). 

\bibitem{peruani08} F. Peruani, A. Deutsch, and M. B{\"a}r, Eur. Phys. J. Special Topics 157, 111 (2008).


\bibitem{baskaran08a} A. Baskaran and M.C. Marchetti, Phys. Rev. E 77, 011920 (2008).	

\bibitem{baskaran08b} A. Baskaran and M.C. Marchetti, Phys. Rev. Lett. 101, 268101 (2008).



\bibitem{vicsek95} T. Vicsek, A. Czirok, E. Ben-Jacob, I. Cohen, and O. Shochet, Phys. Rev. Lett. 75, 1226 (1995).

\bibitem{chate08a} H. Chat{\'e} et al., Eur. Phys. J. B 64, 451 (2008).

\bibitem{chate08}  H. Chat{\'e}, F. Ginelli, G. Gr{\'e}goire, and F. Raynaud, Phys. Rev. E 77, 046113 (2008).

\bibitem{mishra10} S. Mishra, A. Baskaran, and M.C. Marchetti, Phys. Rev. E 81, 061916 (2010).


\bibitem{ginelli10} F. Ginelli, F. Peruani, M. B{\"a}r, and H. Chat{\'e}, Phys. Rev. Lett. 104, 184502  (2010).

\bibitem{smoluchowski1917} M. von Smoluchowski, Z. Phys. Chem. 92, 129 (1917).

\bibitem{huepe04} C. Huepe and M. Aldana, Phys. Rev. Lett. 92, 168701 (2004).

\bibitem{yang08} Y. Yang,  J. Elgeti, and G. Gompper, Phys. Rev. E 78, 061903 (2008).

\bibitem{peruani10} Peruani et al., unpublished (2010).

\bibitem{wensink08} H.H. Wensink and H. L\"owen, Phys. Rev. E 78, 031409 (2008).

\bibitem{escudero10} C. Escudero, F. Maci\'a, and J.J.L. Velazquez, Phys. Rev. E 82, 016113 (2010).



\bibitem{doi} M. Doi and S.F. Edwards, {\it The Theory of Polymer Dynamic} (Clarendon Press, Oxford, 1986).

\bibitem{peruani07} F. Peruani and L.G. Morelli, Phys. Rev. Lett. 99, 010602 (2007).


\bibitem{reif} R. Reif, {\it Fundamentals of Statistical and Thermal Physics} (McGraw-Hill, New-
York, 1965).

\bibitem{dossetti09} V. Dossetti, F.J. Sevilla, and V.M. Kenkre, Phys. Rev. E 79, 051115 (2009).

\bibitem{quintanilla00} J Quintanilla, S Torquato, and R M Ziff, J. Phys. A: Math. Gen. 33, 399 (2000).

\bibitem{schimansky1986} L. Schimansky-Geier, F. Schweitzer, W. Ebeling, and H. Ulbricht, in:
Self-organization by Nonlinear Irreversible Processes, W. Ebeling and
H. Ulbricht, eds. (Springer, Berling, Heidelberg, New York, 1986), p. 67

\bibitem{schweitzer1988a}F. Schweitzer, L. Schimansky-Geier, W. Ebeling, and H. Ulbricht, Physica A
150, 261 (1988).

\bibitem{schweitzer1988b} F. Schweitzer, L. Schimansky-Geier, W. Ebeling, and H. Ulbricht, Physica A
153, 573 (1988).


\bibitem{ramaswamy03} S. Ramaswamy, R.A. Simha and J. Toner, Europhys.Lett. 62, 196 (2003).
























\end{thebibliography}
\end{document}